\documentclass[a4paper]{jpconf}
\usepackage{graphicx}
\usepackage{epsfig}
\usepackage{amssymb}
\usepackage{amsfonts}
\usepackage{amsmath}
\usepackage{euscript}
\usepackage{verbatim}
\usepackage{latexsym}

\newcommand{\beq}{\begin{equation}}
\newcommand{\eeq}{\end{equation}}
\newcommand{\bea}{\begin{eqnarray}}
\newcommand{\eea}{\end{eqnarray}}
\newcommand{\beas}{\begin{eqnarray*}}
	\newcommand{\eeas}{\end{eqnarray*}}

\newcommand{\bquo}{\begin{quote}}
	\newcommand{\enqu}{\end{quote}}
\renewcommand{\(}{\begin{equation}}
\renewcommand{\)}{\end{equation}}

\begin{document}
\title{Gauging away a big bang}

\author{Chethan Krishnan and Avinash Raju}

\address{Centre for High Energy Physics, Indian Institute of Science, Bangalore, India. 560012}

\ead{chethan.krishnan@gmail.com,avinashraju777@gmail.com}

\begin{abstract}
We argue that in the tensionless phase of string theory where the stringy gauge symmetries are unbroken, (at least some) cosmological singularities can be understood as gauge artefacts. We present two conceptually related, but distinct, pieces of evidence: one relying on spacetime and the other on worldsheet.
\end{abstract}

\section{Cosmological Singularities in String Theory}
\noindent
Let us start where we often start: namely, with the observation that an infinite number of independent UV divergences arise when one tries to quantize gravity perturbatively. This infinite number of undetermined couplings means that we need to do an infinite number of experiments to determine them, and this results in a loss of predictivity. This is one reason why we believe that General Relativity must require modifications at short distances.

String theory claims to solve this problem. The way it solves it, is via an enormous gauge symmetry on the worldsheet, called conformal invariance. Even though superficially the theory contains an infinite number of new particles and their couplings, making the situation {\em prima facie} worse, this gauge symmetry is so big that it relates {\em all} of the new couplings to each other. Indeed, one is left with no independent parameters at all in string theory. This makes the theory predictive (in principle) at all scales\footnote{At least after one figures out which solution of the theory corresponds to our Universe! These solutions however can unfortunately be too numerous -- this is the so-called String Landscape with its associated necessity for anthropic arguments.}. This is the jumbo version of a familiar phenomenon from particle physics: the UV breakdown of Fermi theory at the Weak scale requires the introduction of new degrees of freedom, but the new couplings are fixed by the (Higgsed) $SU(2) \times U(1)$ gauge symmetry ensuring that the theory does not give rise to an uncontrollable number of new divergences. 

The worldsheet gauge symmetry of string theory will be an important ingredient in our discussions in this article.

The perturbative non-renormalizability discussed above is a {\em quantum} reason why we believe gravity needs modifications at short distances. But there is also a well-known {\em classical} reason. The singularity theorems of Penrose and Hawking \cite{LSSS} necessitate that in Einstein gravity, regular Cauchy data can evolve and become singular (in the past or future). This is problematic, if we want our theory to be a predictive model of Nature even in the most rudimentary sense. 

One might take consolation in the fact that black hole singularities are covered up by event horizons in classical gravity, at least if Cosmic Censorship holds true\footnote{Of course, the information paradox comes back to bite us at the quantum level, even if this claim was correct!}. But even this grim consolation, at best only holds for an asymptotic observer. An infalling observer still will face loss of predictivity at the future singularity inside the black hole, which he/she is guaranteed to hit. Further, fine-tuned initial data can result in nakedly singular evolution even in classical gravity. Even more relevant for us, the cosmic singularity in our past is naked, and that is something we cannot help but confront. 

Since singularities are believed to be short distance features, they point towards a UV modification of gravity. String theory is expected to be UV-finite, so one might hope that it can help cure singularities. Indeed, stringy resolutions of some singularities do exist \cite{Dixon1, Dixon2, Strominger, APS, McGreevy}, but these are at the level of case studies. It is fair to say that a coherent overall picture of singularities in string theory is lacking.

In particular, cosmological singularities are doubly hard in string theory. This is because cosmology means time-dependence, and quantizing strings in time-dependent backgrounds is difficult. Typically only supersymmetric backgrounds are under control, and supersymmetry forces time {\em  in-}dependence on us.

The question of cosmic singularities in string theory then, at least as we have phrased it above, seems impossible. But one way forward is known. This is to consider timelike/null quotients of Minkowski space as simple examples of singular cosmologies. The tractability of string theory in the covering flat space enables us to probe these singular geometries using strings via orbifold techniques.  Two popular examples where this strategy has been explored are the Milne orbifold and the  null orbifold \cite{HoroSteif, Joan}.

However, all is not well. Tree level string scattering amplitudes on these geometries \cite{LMS1, LMS2, BenMicha, Cornalba, Durin, BenReview} have problematic divergences, so it is unclear what the punchline of this line of work is. The current philosophy \cite{HoroPolch} takes the amplitude at face value and interprets the divergences as due to black hole formation  when the string is inside its own Schwarzschild radius: ie., there is large backreaction from the pointlike ($\alpha'\rightarrow 0$) limit of strings. 

This begs the question: what can be learnt about cosmic singularities from the opposite long-and-floppy limit ($\alpha'\rightarrow \infty$) of strings? This will be the theme of this article, and we will attempt to answer this question from two complementary angles.

\section{Higher Spins and Stringy Gauge Invariance}

Vasiliev has constructed interacting theories involving gravity and higher spins \cite{Vasiliev}.  
There is evidence that these are related to the $\alpha' \rightarrow \infty$ limit of string theory \cite{Sundborg,Witten,Shiraz,Gopakumar}. Higher spin theory (roughly speaking) captures the spacetime dynamics of the worldsheet spectrum of states of the string, in the limit where the string states are massless (masses are inversely related to $\alpha'$).

The presence of extra massless modes suggests that higher spin theory has a lot more gauge invariances than general relativity, which only has diffeomorphisms. Indeed, again loosely, we can think of these higher spin gauge symmetries as the target space realization of worldsheet gauge invariance: this is the un-Higgsed phase of string theory. 

We know that diffeomorphisms (freedom to change coordinates) cannot remove singularities, but these bigger gauge invariances might change that. Could some of the spacetime singularities be just artifacts of a choice of gauge in string theory? Can a higher spin gauge transformation put Milne/Null orbifold in a non-singular gauge? For concreteness, we will consider Milne in this article \cite{KR2}.

\section{Higher Spin Resolution of a Toy Big-Bang} 

We will work with 3D flat space higher spin theory \cite{Arjun,Gonzalez}. AdS gravity in 3D has a tractable Chern-Simons formulation and this can be generalized  \cite{Campoleoni} to higher spins by changing the gauge group from $SL(2){\times}SL(2)$ to $SL(N){\times}SL(N)$. By allowing the gauge fields to take Grassmann values \cite{Avinash} one has a convenient trick\footnote{We call it a trick, but the fact that many things work quite naturally and nicely under the Grassmann approach makes one wonder whether there is more to it than a trick. In particular, general Fefferman-Graham-Banados solutions \cite{Avinash}, asymptotic symmetry algebras \cite{Avinash}, central charges \cite{Avinash}, generalizations to higher spins \cite{Avinash}, addition of chemical potentials \cite{Grumiller} all map conveniently under the Grassmann map. The immediate reason for its success is the fact that $ISO(2,1)$, the gauge group for the Chern-Simons theory that describes flat space gravity is the Inonu-Wigner contraction of $SL(2) \times SL(2)$, the corresponding gauge group in AdS$_3$. } to handle flat space higher spins as well. This enables us to transport much of the AdS technology with little effort to the flat space case. Because the orbifolds we will consider are quotients of 2D flat space, we can always embed them in 3D. 



We take the Milne metric as \cite{Barnich}:
\bea
ds^{2}=-dT^{2}+r_{C}^{2}dX^{2}+\alpha^{2}T^{2}d\phi^{2},\label{eq: Milne metric}
\eea
$X$ is noncompact and $\phi \sim \phi +2 \pi$. 
Spacetime is a double cone with a causal
singularity at $T = 0$ where $\phi$-circle crunches to a point before big-bang-ing. 
This metric is a solution of Einstein equations -- trivially, because it is locally Ricci-flat. We can translate (\ref{eq: Milne metric}) into the $SL(2)$ Chern-Simons language where it corresponds to a flat connection that is invariantly characterized by its holonomy. 

But we can think of Milne also as a solution of the higher spin, ie. $SL(N)$, version of Chern-Simons theory. Now we ask: can one do a gauge transformation involving higher spin charges that preserves the holonomy, while changing the metric so that the singularity is removed? The answer found in \cite{KR2} was that indeed this is possible. It was found that (\ref{eq: Milne metric}) is gauge-equivalent to
\bea
ds^{2}=-dT^{2}+r_{C}^{2}dX^{2}+(\alpha^{2}T^{2}+\beta^2)d\phi^{2},\label{eq: Milne metric res}
\eea
where $\beta^2$ is positive: the crunch-bang singularity is now replaced by a smooth bounce, see figure 1. 
\begin{figure}[h!]
\centering
\includegraphics[width=12cm, height=6cm]{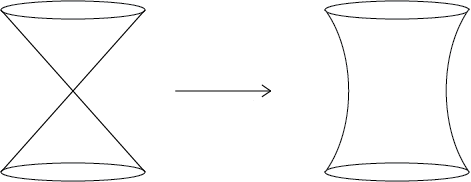}
\caption{}
\end{figure}
In the new gauge there are non-vanishing higher spin fields in the solution which can be thought of as the ``matter" supporting the throat\footnote{This might seem like a violation of singularity theorems, but it is not. Geodesics are no longer good probes in higher spin geometry; the energy conditions (for example) are gauge-{\em variant}. So geodesic incompleteness arguments do not apply.} The curvature invariants as well as the higher spin fields remain finite everywhere. The result in \cite{KR2} was found by turning on spin 3 charges, it was successfully generalized to even higher spins in \cite{Leo}. 
		
Modulo some caveats which we will clarify later, this result is suggestive that the Milne singularity is a gauge artefact.
		
\section{Singularities in the Tensionless String Limit}
Since higher spin theory should capture aspects of the tensionless limit of string theory, our result in the previous section suggests that the divergences in the string amplitudes that people have observed previously should not arise in this limit. The string amplitude, afterall, is a gauge-invariant object. This is what we discuss in this section.
		
In order to investigate the tensionless (ie. $\alpha' \rightarrow \infty$) limit of string amplitudes, we need a dimensionless $\alpha'$. We are working with quotients of flat space, the only available dimensionless $\alpha'$ has to be constructed from the the momenta of the scattering states -- there is no background scale. This is different from the AdS case, where the AdS radius works as a one-parameter family along which we can send the dimensionless $\alpha'$ to $\infty$.

The 4-pt string scattering amplitudes in Milne are complicated beasts even for the lowest lying states, but it turns out that two key results hold \cite{Ben}:
\begin{itemize}
	\item All pathological UV divergences related to the singularity arise only when $\alpha' ({\bf p}_1 -{\bf p}_3)^2 \le 2$ or similar conditions hold. Here ${\bf p_{i}}$ label the momenta of the string scattering states. Roughly speaking the $\alpha'$ is being measured in units of momentum transfer (in $t$ or $u$ channel) and when it is large enough, there are no pathological divergences.
	\item An exhaustive scan reveals that all other divergences are sensible IR divergences. Eg: a whole sequence of logarithmic divergences which have interpretation as the tower of intermediate string states going on-shell.
\end{itemize}
			
This result is re-assuring, because it is a strong suggestion that the resolvability of the Milne singularity we saw in higher spin theory is a reflection of its resolvability in the underlying string theory. 
			
\section{Towards Singularities in the Higgsed Phase of String Theory}
			
Cosmological de Sitter quotients have been resolved previously using higher spins \cite{KR1, Somyadip}. Also, in AdS$_3$ it was noticed that higher spin horizons are gauge-dependent \cite{Kraus,Maloney}. But in both, making a  direct connection to the string worldsheet was hard. But Minkowski orbifolds allow connections both with string theory as well as higher spins. This is the primary source of our interest in them, and the primary reason this work was possible.
			
We focused on Milne in this article, a parallel story holds \cite{Kiran} also for the null Universe of \cite{LMS1}. There are substantial differences in detail and technical difficulty between the two, but satisfyingly, the conclusions are essentially identical.
			
We found a non-singular gauge for the metric, but in these gauges we have non-trivial higher spin fields. These are finite, but typically vanish somewhere in the bulk. Is this bad? We believe this is unlikely, but to answer it unambiguously, without resorting to strings, we will need a higher spin generalization of Riemannian geometry. Since presently we do not have such a theory, we will take the well-definedness of the string amplitude at large $\alpha'$ as an indication that the resolution is legitimate. Also, the fact that we are able to find {\em some} resolution for the metric which preserves the original symmetries of the geometry, we find encouraging. 
			
A consistent set of boundary conditions that contains both the singular (\ref{eq: Milne metric}) and the resolved (\ref{eq: Milne metric res}) solutions was found recently in \cite{Grumiller}. Happily, the gauge transformation that takes the singular geometry to our resolved geometry has zero canonical charge -- ie., they are both the same state, as one would want. This is consistent: one would expect this from the string amplitude argument, because it captures gauge invariant information.
			
Finally, let us note that in standard general relativity one distinguishes between ``coordinate'' singularities and ``true'' singularities. The Schwarzschild horizon in the usual Schwarzschild coordinates is an example of the former: by going to a freely falling (eg: Kruskal) coordinate system at the horizon, one can see that this is an artifact of the Schwarzschild coordinates. ``True'' singularities on the other hand, are places beyond which geodesics cannot be extended even though their affine parameters are finite\footnote{In dimensions four and higher these correspond to regions with infinite curvature. In three dimensions the natural adaptation of this idea is an orbifold singularity, like the one we considered in Milne.}. They cannot be resolved within general relativity and these were our concern in this paper. Our conclusion, at least in the un-Higgsed phase of string theory was that once the stringy gauge symmetries (and not just coordinate transformations) are taken into account, they can also be understood as artifacts of a gauge choice. 
			
The $\alpha' \rightarrow \infty$ limit is precisely the opposite limit of the Einstein gravity limit: strings are long-and-floppy, not pointlike. What we are seeing is an explicit demonstration that the Regge limit can resolve singularities. But the message that  some singularities might be gauge artefacts and might be resolved via gauge transformations is perhaps a useful paradigm to keep in mind, even away from the Regge limit. Can an approach along these lines in a Higgsed phase of string/higher-spin theory shed some light on a more realistic big-bang singularity? 
			
\section*{Acknowledgement}
This article is a summary and synthesis of various papers. We thank Ben Craps, K. Surya Kiran, Shubho Roy, Ayush Saurabh and Joan Simon for related collaborations and Arjun Bagchi, Matthias Gaberdiel, Daniel Grumiller, Shamit Kachru, Elias Kiritsis and Shiraz Minwalla for discussions, comments and/or questions that affected the final presentation. We also thank Matt Lake and Shingo Takeuchi for hospitality at Naresuan University, Phitsanulok, during the IF+YITP symposium, and APCTP for financial support.

\end{document}